\documentstyle[pra,aps,preprint,epsf]{revtex}

\def\be{\begin{equation}}
\def\ee{\end{equation}}
\def\bea{\begin{eqnarray}}
\def\eea{\end{eqnarray}}

\begin{document}
\draft
\begin{titlepage}
\preprint{\vbox{\baselineskip=12pt
\rightline{IP/BBSR/98-35}
\rightline{}
\rightline{hep-th/yymmnn}}}
\title{The Holography Hypothesis and Pre-Big Bang Cosmology}
\author{A. K. Biswas\footnote{e-mail:anindya@iopb.stpbh.soft.net}, 
J. Maharana\footnote{e-mail:maharana@iopb.stpbh.soft.net}
 and R. K. Pradhan\footnote{e-mail:rajat@iopb.stpbh.soft.net}}
\address{Institute Of Physics, Sachivalaya Marg,
 Bhubaneswar-751 005, India}
\date{received}
\maketitle
\begin{abstract}
The consequences of holography hypothesis are investigated for the 
Pre-big-bang string cosmological models. The evolution equations are
obtained from the tree level string effective action. It is shown that 
$\frac{S}{A}$ is bounded by a constant in each case, $S$ being the entropy
within the volume bounded by the horizon of area $A$.
\end{abstract}
\pacs{\tt PACS number(s):~98.80.Cq,~04.30.Db}
\end{titlepage}
The study of the holographic principle has attracted increasing 
attention in the
recent past.
In essence, it states that if we consider a macroscopic theory of space
and everything inside that region, we can represent it by a boundary
theory living on the boundary of that region \cite{hoof,lene}.
In the context of the entropy of the $4-$dimensional black holes, 
it was argued that
all phenomena inside a black hole of size V can be described by
a set of degrees of freedom which live on the surface that bounds V. 
Furthermore, each unit area, in Planck units, of the surface 
contains one bit of information if we imagine
the surface to be a two dimensional lattice. Thus the boundary theory,
in the lattice picture, is discrete and consequently, the information
density is bounded. It has been observed that there is a novel 
correspondence between theories that live in
the bulk and their dual counterparts residing on the boundary through
the holography hypothesis. It has been shown that type IIB string theory
on the background $AdS_5 \times S^5$ and N units of five-form flux on 
$S^5$ is dual to $3+1$ dimensional $U(N)$ super Yang-Mills theory with
16 real supercharges which reside on the boundary of the AdS space \cite{mal}.
There has been considerable activity \cite{gubs,wit,sdwn}
 to investigate various aspects of
the aforementioned duality.\\
 Recently, Fischler and Susskind (FS)\cite{fish} explored the consequences 
of  the holography
principle  in yet another unexplored  and important direction i.e. 
the cosmological domain and derived very  interesting
results. In
the cosmological context, the principle implies that the entropy
contained within a volume of coordinate size $R_{H}$ should not
exceed the area of the horizon in Planck units. Therefore, holography
principle imposes additional constraints on the cosmological models.
First, the consequences of holography were examined
 \cite{fish}, where, the energy density of the Universe is
dominated  by a homogeneous minimally coupled scalar field and later they
studied the Kasner's Universe.\\
The purpose of this note is to examine stringy cosmological models
and the compatibility of such models with holographic principle.
For definiteness, we shall consider the pre-big-bang(PBB) scenario
\cite{vene} which is
endowed with many attractive features. The basic ingredient of the
PBB cosmology is that the Universe started initially
from weak coupling regime with very small curvature. If one assumes
homogeneity to start with, then the Universe undergoes accelerated
expansion due to the fact that the dilaton grows with time in the 
so-called (+)-branch \cite{BV}. 
Thus, a cold, flat,  weakly coupled Universe accelerates and expands 
towards a hot, curved and strongly coupled regime driven by the
dilaton and the  singularity lies in the future. This
super-inflationary growth becomes evident 
 when one works in the string frame metric, 
the metric apprearing in the worldsheet action for a string in the curved
target space. The inflationary solution is related to the expanding 
decelerating
solution, in the post-big-bang era (the singularity is in the past), 
through scale factor duality and
time reversal transformation. In the PBB scenario, we need an exit
from the super-inflationary phase to the standard non-inflationary domain.
Recently, in more general settings, the initial condition of homogeneity
could be relaxed \cite{venez,BMUV} while the Universe evolves from weak 
coupling and low
curvature regime. The Universe proceeds towards the PBB behavior in a suitable 
domain of space and it will fill almost whole space. The Universe appears very 
homogeneous, isotropic and spatially flat within that region.  Furthermore,
in the special case of homogeneous and isotropic cosmologies, which
are not spatially flat, one can obtain explicit solutions\cite{CLW}. 
We refer the
reader to recent review articles on the subject \cite{GVEN} for the current
status \cite{webpage} of PBB cosmology.

In view of the recent attentions on PBB cosmology, it is worthwhile
to study the compatibility of the holographic principle with
various solutions in the PBB scenarios.  

We shall adopt the Einstein frame description 
throughout the course of this work. As is 
well known, the Einstein frame metric 
and  the string frame metric are related 
through a conformal transformation involving the dilaton.  The 
Einstein frame metric has been used to study several interesting aspects of
PBB, especially the scenarios we intend to consider in this note. Furthemore,
the Einstein frame metric is used in deriving  the bounds 
as a consequence of the holography hypothesis. 

We shall compute 
the entropy of the Universe within the horizon as follows: 
first, we determine the entropy per comoving volume, $S^{c}$ using
thermodynamic arguments incorporating the effect of the fluid, namely, 
dilaton and/or axion; and then we obtain the total entropy, $S$,
within the horizon as a product of $S^{c}$ and comoving
volume, $V_{H}^{c}$. The horizon is determined from the 
condition $ds^{2} =0$.

As is well known, when one
considers adiabatically expanding( contracting) Universe, the entropy 
per unit comoving volume remains constant.
We recall \cite{Kolb} that the $0$-th component of the conservation law,
$ T^{\mu}_{\nu;\mu}=0$
 leads to 
\be
\sqrt{g}\frac{dp}{dt} = \frac{d}{dt} (\sqrt{g}(\varrho+p)),
\ee
where, $\varrho$ and $p$ are defined in terms $T^{\mu}_{\mu}$
in the cosmological context and $g$ is the determinant of the
spatial part of the metric.
Then it can be shown that the comoving entropy 
density remains constant in time 
throughout the PBB phase and can 
be expressed as 
\be
S^{c}=\frac{(\varrho+p)\sqrt{g}}{T}
\ee
where,
$T$ is the temperature of the fluid.
Moreover, for $p=\varrho$ 
\be
\varrho = \sigma_{f} T^{2},
\ee
where, $\sigma_{f}$ can be identified to be ``Stefan's constant''
of the fluid and we have set $\hbar=k_{B}=1$ throughout.
We can write $S^{c}$ alternatively as,
\be
S^{c}=2\frac{\sigma^{1/2}}{l_p} (\varrho g)^{1/2},
\ee  
in terms of the dimensionless parameter
$\sigma$, where, $\sigma_{f}=\frac{\sigma}{l_p^{2}}$.

 Here, we have considered the
Universe in PBB regime and assumed that there is no particle production 
during that era.
Let us start with the simplest homogeneous PBB model, viz,
\be
 ds^2 = -dt^2 +\sum(t/t_{0}-1)^{2\lambda_{a}}(dx^{a})^2,
\ee
\be
 \phi(x, t) = \phi_{0} - \sqrt{2} {\sqrt { 1-\sum \lambda _a ^2}}~
 \ln({t\over {t_0}}-1)
\ee
Here, $\{\lambda _a\}$ are independent of x, satisfying 
\be
\sum\lambda_{a}=1,\quad \sum\lambda_{a}^2 = \rho^{2},\quad 
1/3\leq\rho^{2}\leq1
\ee
Note that the first constraint on $\lambda _a$  is the well known Kasner 
condition.
For this case, we get 
\be
V_{H}^{c}= \prod _a(X_{H}^{a}).
\ee
where, the horizon at any instant of time
$t$, is at $X_{H}^{a}$ and is obtained using $ X^{a}(t_{0})=0$
together with the relation 
$ds^2 =0$
The relevant expression  is 
\be
X_{H}^{a}= \frac{t_{0}(\frac{t}{t_{0}} -1)^{1-\lambda_{a}}}{1-\lambda_{a}}.
\ee
The area of the surface bounding that volume  is given by
\be
A_{H}=\big[\prod_a(X_{H}^{a})(\frac{t}{t_{0}} -1)^{\lambda_{a}}]^{2/3}.
\ee
Thus we arrive at 
\be
\frac{S}{A} = \frac{\sigma^{1/2}}{l_p^{2}}\frac{1}{2\sqrt{\pi}}
 \frac{(1-\rho^2)^{1/2}}{\prod(1-\lambda_{a})^{1/3}}
\ee.
It is interesting to point out the similarity of the ratio
 $\frac{S}{A}$ with that
of black holes. We note that if area is measured in $l_p^{2}$ units  
, $\frac{S}{A}$ becomes dimensionless. 
The Kasner condition and the constraint on $\lambda _a ^2$ 
enable us to express
the above ratio  as a function of   
only one $\lambda_{a}$
and let us denote it as  $Y$. Thus 
\be
\frac{S}{A} =\frac{\sigma^{1/2}}{l_{p}^{2}}\frac{1}{2\sqrt{\pi}}
\frac{(1-\rho^2)^{1/2}} {[(1-Y)((Y)^{2} + 
\frac{(1-\rho^2}{2})]^{1/3}}
\ee.
In order to derive an upper(a lower) bound on the above expression we 
need to maximize(minimize)
the right hand side with respect to  $Y$. The desired 
upper(lower) bound on $\frac{S}{A}$ is derived to be
\bea
\frac{S}{A}\leq\frac{\sigma^{1/2}}{l_{p}^{2}}\frac{1}{2\sqrt{\pi}}  
\frac{(1-\rho^2)^{1/2}} {[(11/3 -3\rho^2 -\frac{(3\rho^2 -1)^{3/2}}
{3\sqrt{2}})/9]^{1/3}},\\
\frac{S}{A}\geq\frac{\sigma^{1/2}}{l_{p}^{2}}\frac{1}{2\sqrt{\pi}}  
\frac{(1-\rho^2)^{1/2}}
{[(11/3 -3\rho^2 +\frac{(3\rho^2 -1)^{3/2}}
{3\sqrt{2}})/9]^{1/3}}
\eea
Note the appearance of the constant prefactor 
$\frac{\sigma^{1/2}}{l_{p}^{2}}\frac{1}{2\sqrt{\pi}}   $
and therefore $S/A$ is bounded.
We then obtain  
a bound on $S\over{A}$ over a range of values of
 $\rho ^2$. We have plotted the upper and lower bounds of $r=\frac{S}{A}$
 against $k=\rho^2$ in Figure.1, where the bound
is scaled in the unit of 
$\frac{\sigma^{1/2}}{l_{p}^{2}}$.  
\begin{figure}[h]
\centerline{\leavevmode\epsfysize=5truecm \epsfbox{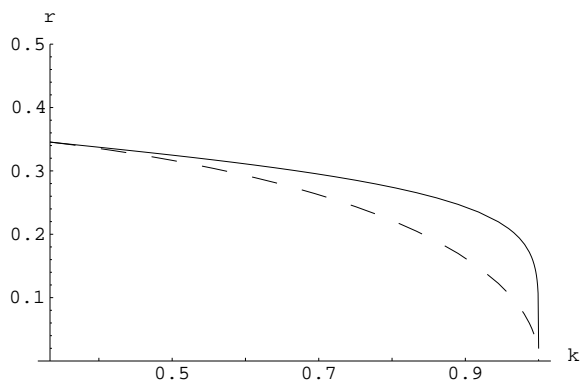}}
\vspace{0.2in}
\caption{ The plot of upper and lower bounds of $r=\frac{S}{A}$ 
vs~~ $k=\rho^{2}$; continuous
and dotted curves are upper and lower bounds respectively.}
\label{a}
\end{figure}

Next we consider a scenario proposed by Maharana, Onofri and 
Veneziano \cite{MOV}.
The starting point is the PBB classical epoch such that the coupling is weak
and curvature is low; therefore, one can trust
 the tree level string
effective action and consequently, the equations of motion are well known. 
Furthermore, the Universe is assumed to be spherically symmetric. It has been 
conjectured [10,11] that a Universe that gives rise to 
dilaton-driven inflation, 
converges in the past, to the Milne metric with constant dilaton. 
In \cite{MOV},
they studied how a small spherically symmetric lump of energy affects the
evolution of the Universe due to classical instability of a perturbed
 Milne metric. As has been noted, for the spherical symmetric 
case, the problem gets simplified considerably when one looks at the Einstein
equations and the matter field equation. It is possible to obtain analytic
asymptotic solutions to the spherically symmetric field equations through
the application of the gradient expansion technique.  
The dilaton , for this case, is given by 

\bea
\phi(\xi,t) = \phi_{0}(\xi) - {2 \over {\sqrt 3}} \sqrt{1-\lambda^2}
\ln(t/(t_{0}(\xi)) -1), \eea

and the line element is
\be
 ds^2 = -dt^2 + (\frac{t}{t_{0}} -1)^{2(1-\lambda)/3}[(\frac{t}
{t_{0}} -1)^{2\lambda}
         e^{2\gamma}d\xi^2 +e^{2\delta}d\omega^2].
\ee 
It looks as if the general solutions given by the above two 
equations depend on four functions of space:
$\lambda , \gamma , \delta $ and $t_0$. We can absorb $\gamma$ in the 
redefinition of the coordinate $\xi$, $dr=e^{\gamma}d\xi$ and by chosing equal
time slices, the spatial dependence of $t_0$ can be removed. Therefore,
one is left with two physically meaningful functions of $\xi$ as should
be the case following the  arguments of \cite{MOV}.
In fact,  we set  $\gamma=\delta=0$ and assume $\lambda$ to be  independent of 
$\xi$ in order to consider  a simple scenario and then  
\be
 ds^2 = -dt^2+(\frac{t}{t_{0}} -1)^{2(1-\lambda)/3}
[(\frac{t}{t_{0}} -1)^{2\lambda}d\xi^2 +d\omega^{2}].
\ee
The position of the horizon $\xi_{H}$ at the time $t$ 
for this asymptotic metric of the Universe is 
\be
\xi_{H}= \frac{3t_{0}}{2(1-\lambda)}(t/t_{0} -1)^{2(1-\lambda)/3} 
\ee
At time $t=t_{0}$ the
horizon is located at $\xi=0$ or the Universe reaches the 
concentrated lump configuration\cite{MOV}.  
Then, the comoving volume is given by
$V_{H}^{c}=\int_{0}^{\xi_{H}}{d\xi}\int{d\omega} = 4\pi\xi_{H}$.
The area of the horizon turns out to be  
\be
A_{H}= 4\pi(\frac{t}{t_{0}} -1)^{\frac{2(1-\lambda)}{3}}.
\ee
Therefore,  
\be
\frac{S}{A} = \frac{\sigma^{1/2}}{l_{p}^{2}}\frac{3}{\sqrt{24\pi}}  
\sqrt{\frac{1+\lambda}{1-\lambda}}.
\ee
So long as $\lambda$ is less than one
$S\over{A}$
is bounded. We may mention in passing that in the numerical 
simulations of \cite{MOV} $\lambda$ turned out to be very small.

It is rather tempting to test whether the generalised 
holographic hypothesis is respected by the  ''Dual Case''.
It is easy to check, for our choice of the parameters,
 $\gamma=0$ and  $\delta=0$, that
the momentum  and Hamiltonian constraints 
are satisfied together with
the rest of the equations of motion under $\lambda \rightarrow -\lambda$ and
$\phi \rightarrow -\phi$.
Thus we can compute  $S\over{A}$ for the ``dual'' case 
and it is given by$\frac{\sigma^{1/2}}{l_{p}^{2}}\frac{3}{\sqrt{24\pi}}  
\sqrt{\frac{1-\lambda}{1+\lambda}}$
 satisfying an upper bound.
 This solution gives post-big-bang branch near 
singularity.

Now we consider the case where the effect of the antisymmetric tensor
field is taken into account besides graviton and the dilaton in the 
four dimensional action \cite{CLW}. Recall that in 
$4$-dimensions the field strength $H_{\mu\nu\lambda}$, through Poincare
duality is expressed as $H_{\mu\nu\lambda} = e^{2\phi}
\epsilon_{\mu\nu\lambda\rho}\partial^
{\rho}h$, where $h$ is the axion. We 
would like to explore whether $\frac{S}{A}$ is bounded or not for the $FRW$ type 
flat metric in the $(+)$ branch. Thus the metric has the form 
$ds^2 = -dt^2 +a(t)^2 dx^{i}dx^{i}$ in the Einstein frame and the 
equations of motion are
\bea
(\dot{a})^{2} + 2a\ddot{a} = -\frac{a^2}{4}
(((\dot{\phi})^{2}+e^{2\phi}(\dot{h})^{2}),\\
a\ddot{\phi} +3 \dot{a}\dot{\phi} - e^{2\phi}a(\dot{h})^{2}=0,\\
\frac{d}{dt}((a^3)e^{2\phi}\dot{h}) =0.
\eea
The Hamiltonian constraint equation is
\be
(\dot{a})^{2} = \frac{a^{2}}{12}((\dot{\phi})^{2}+e^{2\phi}(\dot{h})^{2}).
\ee 
First two equations in the above describe time 
evolution of the 
scale factor and the dilaton. The equation of motion for
the axion is the well known axion 
charge conservation law. We note that dots here denote 
time derivative. It is rather straightforward 
to get the first integral of motion,
\be
H=\pm\frac{1}{\sqrt{12}}\sqrt{(\dot{\phi})^{2}+e^{2\phi}(\dot{h})^{2}}
\ee
where $H$ is the Hubble parameter
and the conservation law yields $\dot{h}=La^{-3}e^{-2\phi} $, $L$ is
chosen to be a positive number. 
Then the comoving volume of the Universe and the area of the horizon are
given respectively by
\be
V_{H}^c= \frac{4}{3}\pi (R_{H})^{3},\quad 
 A=4\pi (a^{2})(R_{H})^{2}
\ee
where, 
as usual, the horizon radius, $R_{H}$, is given by the $ds^2=0$ condition. It 
again turns out that
\be
\frac{S}{A} = \frac{\sigma^{1/2}}{l_p^{2}}\frac{1}{\sqrt{24\pi}}
\ee
where, in the above, $\sigma_{d}+\sigma_{h} =\frac{\sigma}{l_{p}^{2}}$.   
In summary, we have explored the implications of the holographic principle
for several interesting cosmological scenarios in PBB cosmology and 
found that $S\over{A}$ is bounded by constants.
We  mention that the solutions to the metric and dilaton 
considered here follow from the field equations of the tree level 
effective action. 
Thus it is assumed that except the
graviton and dilaton, all other fields which would arise as a consequence
of dimensional reduction of the underlying ten dimensional theory to 
four dimensions
are frozen i.e. carry no spacetime dependence. It is an interesting
issue to envisage the scenario where moduli corresponding to 
internal (compact) dimensions also become time dependent\cite{Lu}. 
 As the dilaton takes large values (strong coupling
domain), the higher derivative terms and higher order stringy correction
effects play an important role and it will be essential to take
into account these effects in string cosmology\cite{poly}.
Again it will be nice derive the
holographic bound for the case when stringy matter is present as
was studied by Gasperini and Veneziano\cite{vene}. 
We hope to present our results in this
direction in a future publication. 
 Moreover, it is an interesting issue to investigate how $S\over{A}$
is bounded at the end of the string phase (i.e. at the begining
of the $FRW$ phase). Veneziano has obtained an interesting relation
\cite{venp}
for the ratio $S\over{A}$ in a general $PBB$ scenario and with
assumptions weaker than Fischler and Susskind \cite{fish}. It will
be interesting to explore the consequences of holography hypothesis
along these lines. 

We would like to thank Gabriele Veneziano for very useful 
correspondence and encouragements. We have benefitted from 
interactions with S. Digal, J. Kamila, R. Roy and S. Sengupta
during the course of this work.

Note added: After completion of this work we became aware of the 
preprint of Dongsu Bak and Soo-Jong-Rey, 
hep-th/9811008, which also discusses holography in string
cosmology.

\end{document}